\documentclass[a4paper,10pt,twocolumn]{article}

\usepackage{amsmath,amsfonts,amssymb,mathtools,amsthm}
\usepackage{bbm} 
\usepackage{epsfig,mathrsfs,enumerate,graphics}
\usepackage[usenames,dvipsnames]{color}
\usepackage{dsfont}
\usepackage{graphicx}
\usepackage{adjustbox}
\usepackage{upgreek}
\usepackage{wrapfig}
\usepackage{caption}
\usepackage{subcaption}
\usepackage{tikz}
\usepackage{minibox}
\usepackage{csquotes}
\usepackage[english]{babel}

\usepackage[style=alphabetic,backend=bibtex8,maxbibnames=10]{biblatex}
\usepackage{datetime2}

\usepackage[all]{xy} \entrymodifiers={!!<0pt,0.7ex>+}

\usepackage[hyperindex=true, 
					hidelinks]{hyperref}

\newcommand{\bg}{\bigskip}

\newcommand{\ens}{\enspace}

\newtheorem{lem}{Lemma}
\newtheorem{thm}[lem]{Theorem}
\newtheorem{prop}[lem]{Proposition}

\theoremstyle{definition}

\newtheorem{Def}[lem]{Definition}

\newcounter{parag}[subsection]

\newcounter{parage}[section]

\newcounter{paraga}

\def\al{\alpha}

\def\ga{\gamma}
\def\Ga{{\Gamma}}
\def\de{\delta}

\def\De{\Delta}
\def\eps{{\varepsilon}}

\def\om{\omega}

\def\Om{\Omega}

\def\sig{{\sigma}}

\def\th{{\theta}}

\newcommand{\ph}{\varphi}

\def\ze{{\zeta}}

\newcommand{\demi}{\frac{1}{2}}

\newcommand{\lan}{\langle}
\newcommand{\ran}{\rangle}

\newcommand{\End}{\operatorname{End}}

\newcommand{\cont}{\operatorname{cont}}

\newcommand{\ID}{\mathop{\hbox{{\rm Id}}}\nolimits}

\newcommand{\I}{{\mathrm i}}
\newcommand{\dd}{{\mathrm d}}

\newcommand{\ee}{\mathrm e}

\newcommand{\pa}{\partial}

\newcommand{\ii}{^{-1}}

\newcommand{\wt}{\widetilde}
\newcommand{\wh}{\widehat}

\newcommand{\RE}{\mathop{\Re e}\nolimits}

\newcommand{\ie}{{\emph{i.e.}}\ }
\newcommand{\cf}{{\it cf.}\ }
\newcommand{\eg}{{\it e.g.}\ }

\newcommand{\rhs}{{right-hand side}}

\newcommand{\BB}{\mathbb{B}}
\newcommand{\C}{\mathbb{C}}

\newcommand{\PPr}{\mathbb{P}}

\newcommand{\Q}{\mathbb{Q}}

\newcommand{\R}{\mathbb{R}}

\newcommand{\SSS}{\mathbb{S}}

\newcommand{\Z}{\mathbb{Z}}

\def\cB{\mathcal{B}}

\newcommand{\Hth}{\mathcal{H}_\th}

\newcommand{\fN}{\mathcal{N}}

\def\cR{\mathcal{R}}

\DeclarePairedDelimiter\abs{\lvert}{\rvert}%
\DeclarePairedDelimiter\norm{\lVert}{\rVert}%

\def\om{\omega}

\newcommand{\ord}{\operatorname{ord}}

\newcommand{\sing}{\operatorname{sing}}

\newcommand{\bem}{\mbox{}^\flat\hspace{-.8pt}}

\newcommand{\ch}[1]{{\stackrel{\raisebox{-.23ex}{$\scriptscriptstyle\vee$}}{#1}}}
\newcommand{\tr}[1]{{\stackrel{\raisebox{-.23ex}{$\scriptscriptstyle\triangledown$}}{#1}}}

\newcommand{\chb}[1]{\raisebox{-.23ex}{${\stackrel{
            \raisebox{-.23ex}{$\scriptscriptstyle\vee$}
          }{#1}
     }$}}

\newcommand{\chn}[1]{\raisebox{.20ex}{${\stackrel{
            \raisebox{-.20ex}{$\scriptscriptstyle\vee$}
          }{#1}
     }$}}
\newcommand{\trn}[1]{\raisebox{.20ex}{${\stackrel{
            \raisebox{-.20ex}{$\scriptscriptstyle\triangledown$}
          }{#1}
     }$}}

\newcommand{\defeq}{\coloneqq} 

\newcommand{\col}{\colon\thinspace}          

\newcommand{\gA}{\mathscr A}       
\newcommand{\gB}{\mathscr B}       
\newcommand{\gD}{\mathscr D}       
\newcommand{\gL}{\mathscr L}       
\newcommand{\gS}{\mathscr S}       
\newcommand{\gU}{\mathscr U}       
\newcommand{\gV}{\mathscr V}       

\newcommand{\chnLth}{\chn\gL\vphantom{\gL}^\th}
\newcommand{\trnLth}{\trn\gL\vphantom{\gL}^\th}

\newcommand{\eith}{\ee^{\I\th}}
\newcommand{\eiths}{\ee^{\I\th_*}}

\newcommand{\est}{{\varnothing}}

\newcommand{\begla}{\begin{equation}}
\newcommand{\beglab}[1]{\begin{equation}	\label{#1}}
\newcommand{\edla}{\end{equation}}

\newcommand{\imp}{\ens\Rightarrow\ens}
\newcommand{\Imp}{\quad\Rightarrow\quad}

\newcommand{\Der}{\operatorname{Der}}

\newcommand{\datestampa}{{\small{File:\ens\hbox{\tt\jobname.tex}
\ens \today}}} 
\newcommand{\datestamp}{{\small{File:\ens\hbox{\tt\jobname.tex}
\ens \DTMnow}}} 

\newcommand{\whRspOm}{\wh\cR\,\!^{\mathrm{s}}_{\Om}}
\newcommand{\tiRspOm}{\wt\cR\,\!^{\mathrm{s}}_{\Om}}
\newcommand{\tiRspdpZ}{\wt\cR\,\!^{\mathrm{s}}_{2\pi\I\Z}}

\newcommand{\whRspZ}{\wh\cR\,\!^{\mathrm{s}}_{\Z}}
\newcommand{\tiRsp}{\wt\cR\,\!^{\mathrm{s}}}
\newcommand{\tiR}{\wt\cR}

\newcommand{\trRspOm}{\raisebox{.1ex}{${\stackrel{
            \raisebox{-.23ex}{$\scriptscriptstyle\bigtriangledown$}
          }{\cR}}$}\,\!^{\mathrm{s}}_{\Om}}
\newcommand{\trR}{\raisebox{.1ex}{${\stackrel{
            \raisebox{-.23ex}{$\scriptscriptstyle\bigtriangledown$}
          }{\cR}}$}}

\newcommand{\euler}{^{\raisebox{-.23ex}{$\scriptstyle\mathrm E$}}}
\newcommand{\stirl}{^{\raisebox{-.23ex}{$\scriptstyle\mathrm S$}}}
\newcommand{\sstirl}{^{\raisebox{-.23ex}{$\scriptscriptstyle\mathrm S$}}}

\newcommand{\Rpos}{\R_{>0}}

\newcommand{\Zp}{\Z_{>0}}
\newcommand{\Znn}{\Z_{\ge0}}

\newcommand{\Ompl}{\Om^+}
\newcommand{\Omn}{\Om^-}

\newcommand{\uOmp}{{\underline{\Om\hspace{.05em}}\hspace{-.05em}'}}

\newcommand{\uN}{{\underline{\fN}}}
\newcommand{\uom}{{\underline{\om}}}
\newcommand{\cuom}{{[\uom]}}
\newcommand{\ua}{{\underline{a}}}
\newcommand{\ub}{{\underline{b}}}

\newcommand{\sh}[3]{\operatorname{sh}\!\big( \begin{smallmatrix}#1,\,
#2\\#3\end{smallmatrix} \big)}

\newcommand{\DD}{%
  {{\Delta}} \hspace{-.0125\textwidth}
  \raisebox{.37ex}{\resizebox{.6ex}{!}{$\mathrm{/}$}}
  \hspace{.007\textwidth} {} 
}

\newcommand{\Gadiff}{\Ga_{\!\textrm{diff}}}
\newcommand{\bb}{\textrm{b}}
\newcommand{\SING}{\operatorname{SING}}
\newcommand{\ANA}{\operatorname{ANA}}

\newcommand{\idm}{{\mathds 1}}

\newcommand{\tcVab}{\gV} 
\newcommand{\tcVae}{\gV^\est} 
\newcommand{\tcVao}{\gV^\uom} 
\newcommand{\tcVaop}{\gV^{\om_1\cdots\om_{r-1}}} 
\newcommand{\hcVao}{\wh\gV^\uom} 
\newcommand{\hcVaop}{\wh\gV^{\om_1\cdots\om_{r-1}}} 
\newcommand{\hcVaun}{\wh\gV^{\om_1}} 

\newcommand{\wc}[1]{\raisebox{-.08ex}{${\stackrel{
            \raisebox{-.23ex}{$\scriptscriptstyle\vee$}
          }{#1}
     }$}}

\newcommand{\ZE}{\operatorname{Ze}}
\newcommand{\WA}{\operatorname{Wa}}

\begin{document}

\thispagestyle{empty}
\begin{center}
{\bf \Large
Resurgence and Mould Calculus}

\bg 

D.~Sauzin,
Capital Normal University, Beijing

on leave from CNRS -- IMCCE, Paris\footnote{\datestampa}

\end{center}

{\small

  \begin{abstract}
  Resurgence Theory and Mould Calculus were invented by J.~\'Ecalle around
  1980 in the context of analytic dynamical systems and are
  increasingly more used in the mathematical physics community, especially since
  the 2010s. We review the mathematical formalism and touch on the applications.
\end{abstract}

 \setcounter{tocdepth}{3}
 \tableofcontents

 }
 

\subsection{Introduction}


Resurgence Theory, founded in the late 1970s by the French
mathematician Jean \'Ecalle in the context of dynamical systems, has
recently become a fixture in the mathematical physics
research literature, with a burst of activity
in applications ranging from
quantum mechanics,
wall-crossing phenomena,
field theory and gauge theory
to string theory.
See \parencite{CW23} for a non-technical account of the importance
taken by Resurgence in current research in Quantum Field Theory.

The theory may be seen as a refinement of the Borel-Laplace summation
method designed to encompass the transseries which naturally arise in
a variety of situations. Typically, a resurgent series is a divergent
power series in one indeterminate that appears as the common
asymptotic expansion to several analytic functions that differ by
exponentially small quantities. The so-called `alien derivations' are
tools designed to handle these exponentially small discrepancies
at the level of the series themselves, through a proper encoding of
the singularities of their Borel transforms.
In the context of local analytic dynamical systems, they have allowed
for quite a concrete description of various moduli spaces.

Together with Resurgence, \'Ecalle also put forward Mould Calculus, a
rich combinatorial environment of Hopf-algebraic nature, designed to
deal with the infinite-dimensional free associative algebras generated
by the alien derivations, but whose scope goes much beyond; for
instance, it provides remarkable tools for the study of the so-called
`multiple zeta values' (MZV).


\subsection{`Simple' version of Resurgence Theory}


Resurgence theory \parencite{Eca81,Eca85} deals with a certain $\C$-algebra $\trR$ over~$\C$ endowed
with a family of $\C$-linear derivations
\begla
\De_\om \in \Der_\C\!\big( \trR \big),
\edla
where~$\om$
runs over the Riemann surface of the logarithm.
For the sake of simplicity, we begin by limiting ourselves to a subalgebra,
$\trRspOm \subset \trR$ \parencite{Eca81,MS16},
easier to describe because it is isomorphic to a subalgebra
$\tiRspOm$ of
formal series in one indeterminate.

\subsubsection{Simple $\Om$-resurgent series}

It is convenient to denote the indeterminate by~$z\ii$ because our formal
series will appear as asymptotic expansions at infinity of functions
analytic in certain unbounded domains.
Let us fix a rank-$1$ lattice~$\Om$ of~$\C$. We shall sometimes use
the notations~`$\om_1$' for a generator of~$\Om$ and
\beglab{eqdefomone}
\om_k \defeq k\,\om_1, \qquad k\in\Z
\edla
for the elements of~$\Om$.

\begin{Def}   \label{defspOmresser}
The space $\tiRspOm$ of \emph{simple $\Om$-resurgent series} is defined as
the set of all formal series $\wt\ph(z) = \sum_{n\ge0} c_n z^{-n} \in \C[[z\ii]]$
such that the Borel transform
\begla
\wh\ph(\ze) = \sum_{n\ge0} c_{n+1} \frac{\ze^n}{n!}
\edla
has positive radius of convergence, defines a
function that has (possibly multivalued) analytic continuation along
any path~$\ga$ starting close enough to~$0$ and avoiding~$\Om$, and all the
branches of the analytic continuation of $\wh\ph(\ze)$ have at worst
`simple singularities'.
\end{Def}

\begin{figure}[h]
  \caption{The path~$\ga$ avoids~$\Om$ and ends at~$\ze_1$,
    near~$\om$.}
  \label{figpathga}
  \begin{center} 
    \includegraphics[width=.407\textwidth]{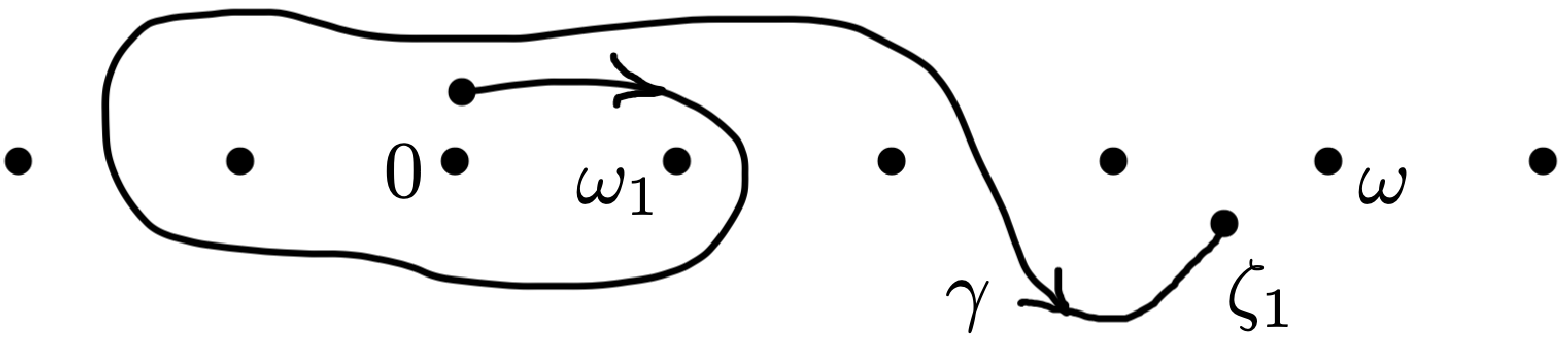}
  \end{center}
  \end{figure}

By `simple singularities', we mean that, if~$\ze_1$ is the endpoint of
a path~$\ga$ as above and we denote by $\cont_\ga\!\wh\ph$ the
analytic continuation of~$\wh\ph$ along~$\ga$, and if~$\om$ is a point of~$\Om$ nearest
to~$\ze_1$ (so that $\cont_\ga\!\wh\ph$ is holomorphic on the
line-segment $[\ze_1,\om[$; see Fig.~\ref{figpathga}), one must have
\beglab{eqsimpsing}
\cont_\ga\!\wh\ph(\om+\xi) = \frac{a_0}{2\pi\I\,\xi} +
\wh\chi(\xi)\frac{\log\xi}{2\pi\I} + F(\xi),
\edla
where $a_0$ is a complex number and both functions $\wh\chi$ and~$F$ extend
analytically through $\xi=0$.

  Defining
  \begla
  \gB \col \C[[z\ii]] \to \C\,\de \oplus \C[[\ze]],
  \qquad
  \gB\wt\ph \defeq c_0 \,\de + \wh\ph
\edla
(here~$\de$ is just a symbol representing the image of~$1$),
we have
  \beglab{eqinclusCVresFormal}
  \C\{z\ii\} \subset
  \tiRspOm = \gB\ii\big( \whRspOm \big) \subset \C[[z\ii]]
\edla
with $\whRspOm \defeq$ the set of all
$c_0\,\de+\wh\ph \in \C\,\de \oplus \C\{\ze\}$ such that $\wh\ph$
extends analytically to the universal cover of $\C-\Om$ and has at
worst simple singularities.

The first inclusion in~\eqref{eqinclusCVresFormal} means that convergent series are
always resurgent: if the radius of convergence of $\wt\ph(z)$ happens to be positive,
then $\wh\ph(\ze)$ extends to an entire function of~$\ze$, thus the
above definition is satisfied with any~$\Om$.

\noindent\emph{Examples of divergent simple $\Om$-resurgent series.}
The Euler series
$\wt\ph\euler(z) \defeq \sum_{p\ge0} (-1)^p p! z^{-p-1}$ (formal
solution to the equation $-\frac{\dd\ph}{\dd z} + \ph=z\ii$) and the
Stirling series
$\wt\ph\stirl(z) \defeq \frac{1}{12}z\ii - \frac{1}{360}z^{-3} +
\frac{1}{1260}z^{-5}+\cdots$ (asymptotic expansion of
$\log\big( \frac{1}{\sqrt{2\pi}} z^{\demi-z}\ee^z \Ga(z)\big)$ as
$\abs z\to\infty$ with $\abs{\arg z} \le\pi-\eps$) have
Borel transforms
\begla
\wh\ph\euler(\ze)= \frac{1}{1+\ze}, \qquad
\wh\ph\stirl(\ze)= \ze^{-2}\Big( \frac{\ze}{2}\coth\frac{\ze}{2} - 1 \Big)
\edla
meromorphic with simple poles in $\Om=\Z$ or $2\pi\I\Z$.
But the typical Borel transforms found in practice are multivalued, as
is the case for instance for $\exp(\wt\ph\stirl(z)) \in \tiRspdpZ$
(asymptotic expansion of $\frac{1}{\sqrt{2\pi}} z^{\demi-z}\ee^z
\Ga(z)$), whose Borel transform can be expressed in terms of the
Lambert~$W$ function \parencite{SauGamma}.
In the latter example, the Borel transform has a principal branch regular at $\ze=0$ but
the other branches of its analytic continuation are singular at~$0$.
A more elementary example of the same phenomenon is offered by
$\operatorname{Li}_2(\ze) \defeq - \int_0^\ze \log(1-\xi)\frac{\dd\xi}{\xi}
\in \whRspZ$.

The growth of the coefficients of a resurgent series $\wt\ph(z)$ is at
most factorial, because $\gB\wt\ph\in\C\,\de\oplus\C\{\ze\}$ implies that
there exist $C,M>0$ such that $\abs{c_n}\le C M^n n!$.
Such series are said to be \emph{$1$-Gevrey}.
If the radius of convergence of $\wt\ph(z)$ is zero, we may hope to get
information on its divergence by studying the
singularities of its Borel transform.
This will be done by means of \'Ecalle's {alien operators} (see below).

\subsubsection{Stability under product and nonlinear operations}

It is obvious that $\tiRspOm$ is a $\C$-vector space stable under
$\frac{\dd\;}{\dd z}$ (note that $\gB$ maps differentiation to
multiplication by $-\ze$).
Much less obviously, it is also stable under multiplication and is thus a subalgebra
of $\C[[z\ii]]$. This is due to the compatibility of the process of
analytic continuation with the convolution
\begla
\wh\ph*\wh\psi(\ze) \defeq \int_0^\ze \wh\ph(u)\wh\psi(\ze-u)\,\dd u
\edla
and to the fact that $\gB$ maps the standard (Cauchy) product of formal series
$\wt\ph\,\wt\psi$ to $\gB\wt\ph*\gB\wt\psi$, where $\de=\gB 1$ is a
unit for the convolution:
one can indeed check that, because~$\Om$ is stable under addition, the
space $\whRspOm$ is stable under convolution
\parencite{Eca81,CNP,stabiconv}.

The algebra $\tiRspOm$ is also stable under nonlinear operations like
substitution into a convergent series, or composition of the form
$\wt\ph\circ (\ID+\wt\psi)$.
For instance, the exponential of any simple $\Om$-resurgent series
and---if it has nonzero constant term---its multiplicative inverse are
simple $\Om$-resurgent series;
also, $\{\ID+\wt\psi\mid \wt\psi \in \tiRspOm\}$ is a group for
composition \parencite{Eca81} (see \parencite{NLresur} for explicit
estimates in the Borel plane that allow one to prove these stability properties).

\subsubsection{Alien operators. Alien derivations}

Notice that in~\eqref{eqsimpsing}, the regular
part~$F$ depends on the choice of the branch of the logarithm,
whereas~$a_0$ and~$\wh\chi$ are uniquely determined.
Thus, for $\ga$ and~$\om$ as above,
we can use the formula
\beglab{eqdefgAomga}
\gA_\om^\ga \wt\ph \defeq
\gB\ii(a_0\,\de + \wh\chi)
\edla
to define a $\C$-linear operator
$\gA_\om^\ga$, which clearly maps $\tiRspOm$ to itself (because $\Om$
is an additive group and~$\wh\chi$, being the difference of two branches of~$\wh\ph$ translated
by~$\om$, must belong to $\whRspOm$).

We call \emph{alien operators} of $\tiRspOm$ the elements of the
subalgebra of $\End_\C\!\big( \tiRspOm \big)$ generated by the $\gA_\om^\ga$'s.
Note that they all annihilate convergent series, since $\C\{z\ii\}\subset\ker(\gA_\om^\ga)$.
Among them, we single out two families of operators,
$(\De_\om^+)_{\om\in \Ompl}$ and $(\De_\om)_{\om\in \Ompl}$, where
\beglab{eqdefomk}
\Ompl \defeq \{ \om_k \mid k\in\Zp \}
\edla
with reference to~\eqref{eqdefomone}.

\begin{figure}[h]
\caption{An example of path $\ga(\eps)$, here with $r=4$ and $\eps=(+,-,+)$.}
\label{figpathgaeps}
\begin{center} 
    \includegraphics[width=.443\textwidth]{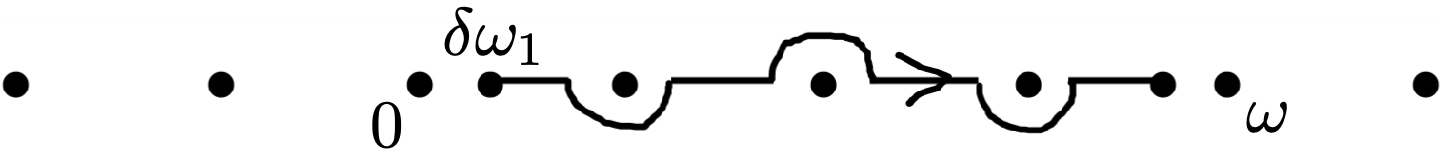}
\end{center}
\end{figure}

\begin{Def}   \label{defDepDe}
Let $\om=\om_r\in \Ompl$ ($r\in\Zp$).
We consider, for each
$\eps=(\eps_1,\ldots,\eps_{r-1})\in\{+,-\}^{r-1}$, the path
$\ga(\eps)$ that connects $\de\om_1\in \, ]0,\om_1[$, where
$\de\in(0,\demi)$, to $(r-\de)\om_1\in \, ]\om_{r-1},\om[$ by
following $]0,\om[$ except that $\om_k$ is circumvented to the right
(resp.\ left) if $\eps_k=+$ (resp.\ $-$). See Fig.~\ref{figpathgaeps}.
We set
\beglab{eqdefDepDe}
  \De^+_\om \defeq \gA_\om^{\ga(+,\ldots,+)},
    \ens
  \De_\om \defeq \sum_{\eps\in\{+,-\}^{r-1}}
  \frac{p(\eps)!q(\eps)!}{r!} \gA_\om^{\ga(\eps)},
  \edla
  where $p(\eps)$ and $q(\eps)=r-1-p(\eps)$ are the numbers of symbols~`$+$'
  and~`$-$' in the tuple~$\eps$.
\end{Def}

The weights used in~\eqref{eqdefDepDe} for the second family have been
chosen so that these operators satisfy the Leibniz product rule, for
which reason $\De_\om$ is called the \emph{alien derivation with
index~$\om$}.

\begin{thm}[\parencite{Eca81}]   \label{thmDeomderiv}
For each $\om\in\Ompl$, the operator $\De_\om$ is a
derivation of the algebra $\tiRspOm$.
  \end{thm}

There is even an `alien chain rule'. For instance, for the Stirling
series,
$\De_\om \big(\ee^{\wt\ph\sstirl}\big) =
(\De_\om\wt\ph\stirl)\ee^{\wt\ph\sstirl}$,
with $\De_\om\wt\ph\stirl= 1/r$ if $\om=2\pi\I r$.

One way of proving Theorem~\ref{thmDeomderiv} is to check that one can
express~$\De_{\om_r}$ in terms of $\De^+_{\om_1},\ldots,\De^+_{\om_r}$
and that each~$\De^+_\om$ satisfies a modified Leibniz rule:
\beglab{eqmodifLeibniz}
\De^+_{\om_r} (\wt\ph\,\wt\psi) =
(\De^+_{\om_r}\wt\ph) \, \wt\psi
+ \sum_{k=1}^{r-1} (\De^+_{\om_{r-k}}\wt\ph) \De^+_{\om_k}\, \wt\psi
+ \wt\ph \De^+_{\om_r}\, \wt\psi.
\edla
A more conceptual explanation stems from the relation of these operators to the 
symbolic Stokes automorphism described in Section~\ref{secTrsSa} below.

Let $\Omn \defeq \{ -k\om_1 \mid k\in\Zp \}$. By repeating the previous
definitions with $\om_1$ changed into $-\om_1$, we get alien operators
$\De^+_\om$ and alien derivations $\De_\om$ for all $\om\in \Om' \defeq \Om-\{0\}$.

\noindent \emph{Large~$n$ asymptotics of the $c_n$'s.}
Notice that, for a given $\wt\ph(z)=\sum_{n\ge0}c_nz^{-n}\in\tiRspOm$,
since the $c_n$'s appear as coefficients in the Taylor expansion of
$\wh\ph(\ze)$ which has $\pm\om_1$ as its nearest-to-origin potential
singular points, their growth is dictated by the nature of the
singularities there, \ie by $\De_{\om_{\pm1}}\wt\ph =
\De^+_{\om_{\pm1}}\wt\ph$. For instance, denoting by~$a_\pm$ the
constant term in $\De^+_{\om_{\pm1}}\wt\ph$, one can prove that
\begla
c_{n+1} = -\frac{1}{2\pi\I} n! \big(\om_1^{-n-1}a_++ \om_{-1}^{-n-1}a_- +
O(\abs{\om_1}^{-n}n\ii) \big)
\edla
and one can refine this estimate to arbitrary precision by using later
terms in $\De^+_{\pm\om_1}\wt\ph$, and even get a transasymptotic
expansion (involving corrections of orders
$\om_{\pm2}^{-n},\om_{\pm3}^{-n},\ldots$) by incorporating
contributions from
$\De^+_{\pm\om_2}\wt\ph, \De^+_{\pm\om_3}\wt\ph,\ldots$

\subsubsection{Freeness of the alien derivations and a first glimpse
  of mould calculus}

\label{secFree}

The term `alien derivation' was coined by \textcite{Eca81} to highlight how this
array of operators is different in nature from the natural derivation $\frac{\dd\;}{\dd z}$.
If we focus on $\Der_\C(\tiRspOm)$, viewed as a Lie subalgebra of
$\End_\C(\tiRspOm)$ with commutator as Lie bracket,
we have the relation
\begla
\big[ \tfrac{\dd\;}{\dd z}, \De_\om \big] = \om \De_\om
\quad\text{for each $\om\in\Om'$}
\edla
(easy to check from the definition),
but the striking fact is that the Lie subalgebra generated by
$(\De_\om)_{\om\in\Om'}$ under Lie bracketing and multiplication by arbitrary
elements of $\tiRspOm$ is isomorphic to the tensor product of
$\tiRspOm$ with the free Lie algebra over $\Om'= \Om-\{0\}$.
We thus have, in this very analytic context, an infinite-dimensional
free Lie algebra!
This is in marked contrast with $\C[[z\ii]]$, all of whose derivations
are generated by $\tfrac{\dd\;}{\dd z}$, \ie of the form
$\al(z)\tfrac{\dd\;}{\dd z}$, $\al\in\C[[z\ii]]$.

We can also consider the associative subalgebra of $\End_\C(\tiRspOm)$
generated by the $\De_\om$'s, and there is a similar statement.
Formulating precisely these facts will usher us into the realm of mould
calculus.

From now on we relinquish the notations~`$\om_1$' and~`$\om_k$'
of~\eqref{eqdefomone} and use $\om_1,\ldots,\om_r$ to denote generic
elements of~$\Om'$.
In fact, we will view~$\Om'$ as an alphabet, whose
letters form words $\uom=\om_1\cdots\om_r$ (with arbitrary
$\om_1,\ldots,\om_r\in\Om'$ and $r\in\Znn$, including the empty
word~$\est$ in the case $r=0$).
We denote by~$\uOmp$ the set of all words on~$\Om'$.

\begin{thm}[\parencite{Eca81}]  \label{thmfreeness}
  Let
  \begla
  \De_\uom \defeq \De_{\om_r}\cdots \De_{\om_1}
  \ens\text{for any $\uom=\om_1\cdots\om_r\in\uOmp$,}
  \edla
  with the convention $\De_\est\defeq \ID$.
Consider, for any finitely-supported family
  $\Phi=(\Phi^{\uom})_{\uom\in\uOmp}$ of elements of $\tiRspOm$, 
the operator
\beglab{eqdefPhiDD}
\Phi\DD \defeq \sum_{\uom\in\uOmp} \Phi^\uom \De_\uom
\in \End_\C(\tiRspOm).
\edla
Then $\Phi\DD$ is nonzero, unless all coefficients $\Phi^\uom$ are zero.
  \end{thm}

  For instance, $\De_{\om_1}\De_{\om_2}$ is a non-trivial operator for
  any $\om_1,\om_2\in\Om'$, and
  $\De_{\om_1}\De_{\om_2} - \De_{\om_2}\De_{\om_1}$ is a non-trivial
  derivation if $\om_1\neq\om_2$.

  In a formula like~\eqref{eqdefPhiDD}, the family
  $\Phi=(\Phi^{\uom})_{\uom\in\uOmp}$ is called an $\tiRspOm$-valued
  \emph{mould}, and the \rhs\ is called a \emph{mould expansion}---see 
  Section~\ref{secMC} on Mould Calculus below.
  Here, the finite support condition is needed to make
  sense of the summation, but later in Section~\ref{secMC}
  we will consider
  moulds that are not necessarily finitely-supported and take their
  values in an arbitrary commutative algebra, not necessarily~$\tiRspOm$.

  The proof of Theorem~\ref{thmfreeness} follows from
  Theorem~\ref{thmexistgUreinf} given in Section~\ref{secMC}.
    The idea is that, according to Theorem~\ref{thmexistgUreinf},
    there is an $\tiRspOm$-valued mould
    $(\gU^{\uom})_{\uom\in\uOmp}$ such that $\gU^\est=1$ and, for every
    $\om_0\in\Om'$ and $\uom=\om_1\cdots\om_r\in\uOmp$,
    \begla
    \De_{\om_0}\gU^{\om_1\cdots\om_r} = 
      \begin{cases}
        \gU^{\om_2\cdots\om_r} & \text{if $r\ge1$ and $\om_0=\om_1$}
        \\[1ex]
        0 & \text{otherwise}.
      \end{cases}
      \edla
Notice then that the support of $(\gU^{\uom})_{\uom\in\uOmp}$ must be all
of~$\uOmp$ because
$\De_{\om_r}\cdots\De_{\om_1}\gU^{\om_1\cdots\om_r}=\gU^\est\neq0$. More
generally, for any $\ua\in\uOmp$,
\begla
\De_\ua \gU^{\uom} =
      \begin{cases}
        0 & \text{if $\ua$ is not a prefix of $\uom$}
        \\[1ex]
        \gU^\ub & \text{if $\uom=\ua\, \ub$},
      \end{cases}
\edla
which easily implies Theorem~\ref{thmfreeness} since it gives
$(\Phi\DD)\gU^{\uom^*} = \Phi^{\uom^*}$ for any~$\uom^*$ of minimal
length in the support of~$\Phi$.

%
    A mould~$\Phi$ is called \emph{alternal} if
    \beglab{eqaltPhi}
    \Phi^\est=0 \ens\text{and}\ens
    \sum_{\uom\in\uOmp} \sh{\ua}{\ub}{\uom} \Phi^{\uom} = 0
    \;\text{for any}\ \ua,\ub\neq\est,
    \edla
    where the `shuffling' coefficient $\sh{\ua}{\ub}{\uom}$ counts the number of
    permutations that allow to interdigitate the letters of~$\ua$
and~$\ub$ so as to obtain~$\uom$ while preserving their internal order
in~$\ua$ or~$\ub$. For instance, the condition~\eqref{eqaltPhi} with
$\ua=\om_1$ and $\ub=\om_2$ says that
$\Phi^{\om_1\om_2}+\Phi^{\om_2\om_1}=0$ and, with $\ua=\om_1$ and
$\ub=\om_1\om_2$, that
$2\Phi^{\om_1\om_1\om_2}+\Phi^{\om_1\om_2\om_1}=0$.

We shall say more on alternality and related mould symmetries in
Section~\ref{secMC}, but let us already mention that one interest of
alternal moulds lies in the following
\begin{prop}
For any alternal $\tiRspOm$-valued mould $(\Phi^\uom)_{\uom\in\uOmp}$, the
operator~\eqref{eqdefPhiDD} satisfies
\beglab{eqPhiDDalt}
\Phi\DD  = \sum_{\uom\in\uOmp} \Phi^\uom \De_\uom =
\sum_{\uom\in\uOmp-\{\est\}} \tfrac{1}{r(\uom)} \Phi^\uom \De_\cuom
\edla
where $r(\uom)$ denotes the length of the word~$\uom$ and
\begla
\De_\cuom \defeq [\De_{\om_r},[\ldots[\De_{\om_{2}},\De_{\om_1}]\ldots]]
\quad\text{for $\uom=\om_1\cdots\om_r$}
\edla
(with the convention $\De_{[\om_1]}=\De_{\om_1}$ if $r=1$).
In particular, this alien operator is a derivation:
\begla
\text{$\Phi$ alternal} \Imp
\Phi\DD=\sum_{\uom\in\uOmp} \Phi^\uom \De_\uom
  \in \Der_\C(\tiRspOm).
\edla
\end{prop}

Thus, in view of Theorem~\ref{thmfreeness}, when dealing with
operators of the form~\eqref{eqPhiDDalt}, \ie in the alternal case,
we are actually working in the free Lie algebra on~$\Om'$ (but we use
an indexation by all the words and restrict to coefficients given by
alternal moulds, instead of using a Lyndon basis or a Hall basis).

\subsection{Simple resurgent functions obtained by Borel-Laplace
  summation}

\subsubsection{Borel-Laplace summation}

Many resurgent series encountered in practice are useful because they
are also `summable' (or `accelero-summable', but we won't touch on
accelero-summability in this article).

Borel-Laplace summation is the application of Laplace transform (or
some variant) to $(\gB\wt\ph)(\ze)$, giving rise to a
function~$\ph(z)$ for which the formal series~$\wt\ph(z)$ appears as
asymptotic expansion at infinity;
but the function~$\ph$ is usually not analytic in a full
neighbourhood of infinity, only in a sectorial neighbourhood,
because~$\wt\ph$ is usually a divergent series.

\begin{figure}[h]
  \caption{Above: Directions for Laplace integration with
    $J=(\th_1,\th_2)$. Below: the  union of half-planes~$\gD^J$.}
  \label{figdomgDJ}
  \begin{center} 
    \includegraphics[width=.36\textwidth]{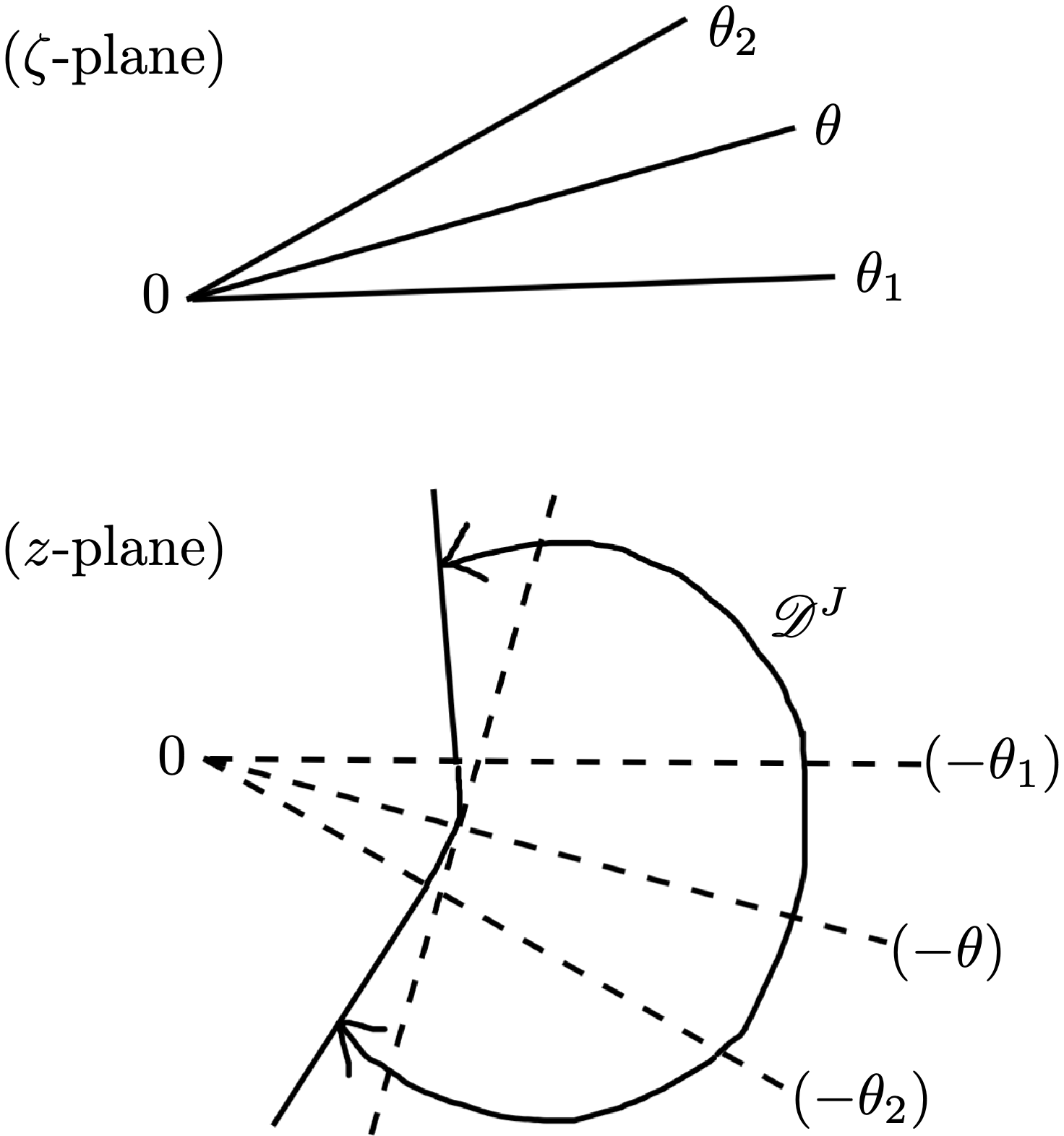}
  \end{center}
\end{figure}

A condition is needed for this to be possible. Let $J$ 
denote a real interval. We say that $\wt\ph(z)$
is $1$-summable in the directions of~$J$ if the function $\wh\ph(\ze)$
extends analytically to the sector $\{\arg\ze\in J\}$ and satisfies there
an exponential bound,
$\abs{\wh\ph(\ze)} \le C\,\ee^{c\abs\ze}$ for some $c,C>0$.
Each of the Laplace transforms
\beglab{eqdefLaplth}
(\gL^\th\!\gB\wt\ph)(z) \defeq c_0 + \int_0^{\eith\infty}
\ee^{-z\ze}\,\wh\ph(\ze)\,\dd\ze,
\qquad \th\in J
\edla
is then analytic in the half-plane $\{\RE(z\eith)>c\}$. By the Cauchy
theorem, these functions match and can be glued so as to define one
function $(\gL^J\!\gB\wt\ph)(z)$ analytic in the union~$\gD^J$ of these
half-planes---see Figure~\ref{figdomgDJ}.
The domain~$\gD^J$ can be viewed as a sectorial neighbourhood of
infinity of opening $\abs{J}+\pi$ (to be considered as a part of the
Riemann surface of the logarithm if $\abs{J}>\pi$).

In that situation, the function $\gS^J\wt\ph \defeq \gL^J\!\gB\wt\ph$
satisfies the asymptotic property $(\gS^J\wt\ph)(z) \sim
\wt\ph(z)$ as $\abs z\to\infty$ with $1$-Gevrey qualification---a
classical fact that sometimes goes under the name of Watson's
lemma, see \eg \parencite{Cos09,MS16}.
The operator
\begla
\gS^J = \gL^J\circ\gB
\edla
is thus called the Borel-Laplace
summation operator in the directions of~$J$.
One can easily check that
\beglab{eqdiffdiffSJ}
\gS^J \tfrac{\dd\wt\ph}{\dd z} = \tfrac{\dd\;}{\dd z} (\gS^J \wt\ph)
\;\text{and}\;
\gS^J\!\big[ \wt\ph(z+\al) \big] = (\gS^J \wt\ph)(z+\al)
\edla
for any $\al\in\C$.
The summation operator is also an algebra homomorphism. Indeed, $\gB$
maps the Cauchy product of formal series to convolution in
$\C\,\de\oplus\C\{\ze\}$, which~$\gL^J$ maps to the pointwise product
of functions:
\beglab{eqmultSJ}
\gS^J( \wt\ph\, \wt\psi) = (\gS^J\wt\ph) (\gS^J\wt\psi). 
\edla

Equations~\eqref{eqdiffdiffSJ}--\eqref{eqmultSJ} show that, if we
start with~$\wt\ph(z)$ formal solution to an ordinary differential
equation or a difference equation, even a nonlinear one, and if we can
subject it to Borel-Laplace summation, then $\gS^J\wt\ph(z)$ is an
analytic solution to the same equation.
Notice however that there may exist other solutions with the same
asymptotic series~$\wt\ph$, maybe obtained by means of $\gS^{J'}$ with
a different interval~$J'$, or by some other means...

The Borel-Laplace picture becomes particularly interesting when we
start with a resurgent series $\wt\ph(z)$, in which case
$\ph^J(z)\defeq \gS^J\wt\ph(z)$ deserves to be called a
\emph{resurgent function}.

\begin{center}
\begin{picture}(6,3.8)
\thinlines

\put(1.05,.45){\makebox(0,0)[br]{Resurgent\ }}
\put(1.05,.01){\makebox(0,0)[br]{series $\wt\ph(z)$}}
\put(4.4,2.05){\makebox(0,0)[l]{$c_0\de+\wh\ph(\ze)$ with }}
\put(4.4,1.65){\makebox(0,0)[l]{good analytic}}
\put(4.4,1.25){\makebox(0,0)[l]{continuation}}
\put(1.3,3.6){\makebox(0,0)[tr]{Resurgent\; \, }}
\put(1.3,3.2){\makebox(0,0)[tr]{function $\ph^J(z)$}}

\put(1.5,.6){\vector(3,1){2.5}}
\put(4,2){\vector(-3,1){2.5}}

\put(2.6,1.1){\makebox(0,0)[br]{Borel $\gB$}}
\put(2.65,2.6){\makebox(0,0)[bl]{Laplace $\gL^J$}}

\multiput(.3,.86)(0,.34){5}{\line(0,1){.185}}
\put(.3,2.47){\vector(0,1){.2}}

\end{picture}
\end{center}

\subsubsection{Transseries, symbolic Stokes automorphism}   \label{secTrsSa}

\begin{figure}[h]
  \caption{Decomposition of the contour for the difference of two
    Laplace transforms as a sum of contours.}
\label{figpathdiff}
\begin{center} 
    \includegraphics[width=.438\textwidth]{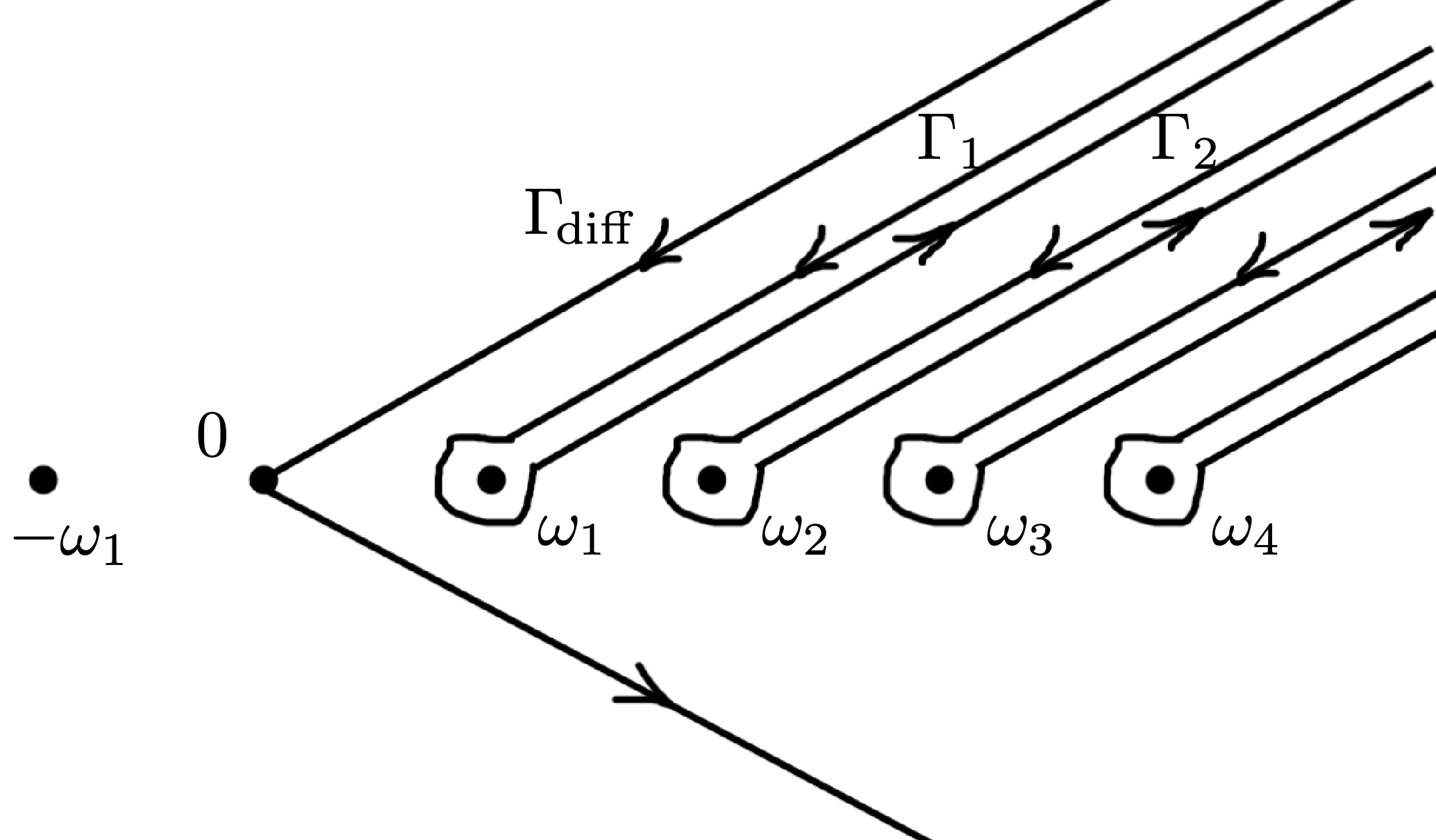}
\end{center}
\end{figure}

Indeed, let us pick a generator~$\om_1$ of~$\Om$ and $\th_*\in\R$ so that
$\Ompl \subset d\defeq\eiths \Rpos$ (making use of
notations~\eqref{eqdefomone} and~\eqref{eqdefomk}).
Suppose that $\wt\ph \in \tiRspOm$ is
$1$-summable in the directions of $J_+\defeq (\th_*-\de,\th_*)$ as
well as those of $J_-\defeq (\th_*,\th_*+\de)$ for some $\de\in(0,\pi)$.
We then have two Borel sums at our disposal, $\gS^{J_+}\wt\ph(z)$ and
$\gS^{J_-}\wt\ph(z)$. Comparing them in the domain
$\gD^{J_+} \cap \gD^{J_-} \cap \{\RE(z\eiths)>0\}$, we find
\begla
\gS^{J_+}\wt\ph(z) -\gS^{J_-}\wt\ph(z) = \int_{\Gadiff}\!
\ee^{-z\ze}\,\wh\ph(\ze)\,\dd\ze
\edla
with a contour~$\Gadiff$ that can be decomposed in a sum of contours
$\Ga_1,\Ga_2,\ldots$ as illustrated on Figure~\eqref{figpathdiff}.

By the Cauchy theorem we thus get
\beglab{eqdiffgsJpm}
\gS^{J_+}\wt\ph(z) = \gS^{J_-}\wt\ph(z) +
\sum_{k\ge1} \int_{\Ga_k}
\ee^{-z\ze}\,\wh\ph(\ze)\,\dd\ze,
\edla
at least if the growth of the relevant branches of~$\wh\ph(\ze)$ along
the~$\Ga_k$'s stays at most exponential and the series over~$k$ is
convergent (if not, \eqref{eqdiffgsJpm} still has asymptotic meaning upon
appropriate truncations).
The point is that, for each $k\ge1$,
in view of \eqref{eqsimpsing}, \eqref{eqdefgAomga} and~\eqref{eqdefDepDe}, the branch
of~$\wh\ph(\ze)$ that we are using along~$\Ga_k$ has a pole and a
monodromy variation precisely described by $\gB \De^+_{\om_k}\wt\ph$, hence
\begla
\int_{\Ga_k} \ee^{-z\ze}\,\wh\ph(\ze)\,\dd\ze =
\ee^{-\om_k z} \gS^{J_-} \De^+_{\om_k}\wt\ph(z)
\edla
(use the change of variable $\ze=\om_k+\xi$ and
\eqref{eqfirstHklLapl}--\eqref{eqdefchb} below) and, finally,
\beglab{eqdiffgSpmfinal} 
\gS^{J_+}\wt\ph = \gS^{J_-}\wt\ph +
\sum_{k\ge1} \ee^{-\om_k z} \gS^{J_-} \De^+_{\om_k}\wt\ph.
\edla

The \rhs\ of~\eqref{eqdiffgSpmfinal} can be interpreted as the
Borel-Laplace summation of a transseries, \ie we are naturally led to
work with expressions of the form
\begla
\wt\Psi=\wt\psi_0 + \sum_{k\ge1} \ee^{-\om_k z} \wt\psi_k,
\qquad \wt\psi_0, \wt\psi_1, \wt\psi_2\ldots\in\tiRspOm,
\edla
which live in the space
\begla
\tiRspOm[[\ee^{-\om_1 z}]] = \underset{k\ge0}{\wh\oplus} \ee^{-k\om_1 z} \tiRspOm.
\edla
We view that space as a completed graded algebra,
in which acts the operator
\beglab{eqdefDDpd}
\DD^+_d \defeq \ID + \sum_{\om\in \Om\cap d} \ee^{-\om z} \De^+_{\om},
\edla
and to which we extend~$\gS^{J_\pm}$ by declaring that
$\gS^J[\ee^{-\om z}\wt\Psi] \defeq \ee^{-\om z}\gS^J\wt\Psi$,
the upshot being
\beglab{eqSJpSJmDDp}
\gS^{J_+} = \gS^{J_-} \circ \DD^+_d,
\edla
at least in restriction to those transseries of $\tiRspOm[[\ee^{-\om_1
  z}]]$ all of whose components satisfy the growth and convergence
requirements needed for~\eqref{eqdiffgsJpm}.

We can now reach a more conceptual understanding of
Theorem~\ref{thmDeomderiv} and the identities~\eqref{eqmodifLeibniz}:
\begin{itemize}
\item
  The identities~\eqref{eqmodifLeibniz} express that~$\DD^+_d$ is an
  algebra endomorphism, which is no surprise since~\eqref{eqSJpSJmDDp}
  involves two algebra homomorphisms $\gS^{J_\pm}$.
\item
  The formulas expressing~$\De_{\om_r}$ in terms of
  $\De^+_{\om_1},\ldots,\De^+_{\om_r}$ alluded to right after the
  statement of Theorem~\ref{thmDeomderiv} are just the
  homogeneous components of the relation
  \beglab{eqlogDDpd}
  \log\DD^+_d = \sum_{\om\in \Om\cap d} \ee^{-\om z} \De_{\om},
  \edla
  which is the true origin of formula~\eqref{eqdefDepDe}.  
\item
  In the context of a completed graded algebra like ours, the logarithm of an
  algebra automorphism is always a derivation, and the homogeneous
  components of a derivation are all derivations,
  this explains why each~$\De_\om$ is a derivation of $\tiRspOm$.
\end{itemize}

\begin{Def}
  The operator~$\DD^+_d$ is called the symbolic Stokes automorphism in
  the direction $d=\eiths \Rpos$.
  The operator $\DD_d\defeq \log\DD^+_d$ is called the directional alien
  derivation in the direction~$d$.
  \end{Def}

Here we see how we can go beyond the traditional theory of asymptotic series, in
which a function is determined by its asymptotics only up to an
infinitely flat function.
In the framework of $1$-Gevrey asymptotics, flat functions are in fact
exponentially small, and the previous computation offers a glimpse of
how the resurgent tools gives us a handle on the exponentially small
ambiguities inherent to the situation.

The point is that~$\DD^+_d$ is an algebra automorphism that commutes
with $\frac{\dd\;}{\dd z}$ and composition with $z\mapsto z+\al$,
and thus preserves the property of being a formal solution to a
differential or difference equation.
One can construct other such automorphisms;
the simplest example is $(\DD^+_d)^w$, with arbitrary $w\in\C$, which
can be written as a mould expansion:
\begla
(\DD^+_d)^w = \sum_{\uom\in\uOmp} \tfrac{w^{r(\uom)}}{r(\uom)!}
\ee^{-\norm\uom z}\De_\uom
\edla
with the notation $\norm{\om_1\cdots\om_r}\defeq
\om_1+\cdots+\om_r$.
The case $w=\pm\demi$ gives rise to the median summation operator,
$\gS^{\th_*}_{\operatorname{med}} \defeq \gS^{J_+} \circ
(\DD^+_d)^{-\demi} = \gS^{J_-} \circ (\DD^+_d)^{\demi}$,
especially useful when $\th_*=0$ and real symmetry is at
play.\footnote{%
  Lack of space prevents us from touching here upon \'Ecalle's
  `well-behaved real-preserving averages', which offer an alternative
  approach to real summation---the reader may consult \parencite{EM96}
  and \parencite{MeGUA}.
  }
 
All this can be particularly relevant in physics, where one often starts by
developing a so-called perturbative theory, where various formal
expansions naturally appear; but then, if they can be subjected to the
resurgent apparatus, one may hope to incarnate them as functions with
meaningful exponentially small contributions related to
non-perturbative physics...
This may be viewed as one striking success of Resurgence Theory:
to give systematic mathematical tools allowing one to extract non-perturbative
contributions from a perturbative divergent series, via the analysis
of the singularities in the Borel plane,
in line with the expectations of, for instance, \textcite{Par78} and
't~Hooft \parencite{tHooft79}.
  
\subsection{The algebra $\trR$ of general resurgent singularities}

We now relax one by one the two requirements imposed by
Definition~\ref{defspOmresser}, namely that the only obstacles to
analytic continuation may occur at points of a given lattice~$\Om$ and
that the singularities are no worse than simple.

\subsubsection{Relaxing the constraint on the location of singularities} 

\begin{Def}
  \label{defendlesscont}
  We call \emph{endlessly continuable} any~$\wh\ph(\ze)$ analytic near~$0$
  for which, for every real $L>0$, there exists a finite subset~$F_L$
  of~$\C$ such that~$\wh\ph(\ze)$ can be analytically continued along
  every Lipschitz path of length~$<L$ starting in the initial domain
  of definition of~$\wh\ph$ and avoiding~$F_L$.
\end{Def}

Variants are possible---see \parencite{CNP,KS20}, or \parencite{Eca85} for the most
general definition (`continuable with no cut').
The point is that singular points are possibly very numerous but,
in a sense, still isolated.
This leaves room for a behaviour such as the one envisioned by
\textcite{Vor83}: there may exist a dense subset~$\Om$ of~$\C$ such
that, at every point of~$\Om$, at least one of the branches of the
analytic continuation of~$\wh\ph$ is singular
(but, in any given region of~$\C$, one `sees' only finitely many
singularities at a time: you need longer and longer paths of analytic
continuation to see more and more singularities in that region).

  %
Denoting by $\tiRsp$ the space of all
  $\wt\ph(z)\in\gB\ii(\C\,\de\oplus\C\{\ze\})$ for which $\wh\ph(\ze)$
  is endlessly continuable with at worst simple singularities for all
  the branches of its analytic continuation,
  %
  %
  we can now define the alien operators
  \begla
  \De^+_\om \in \End_\C(\tiRsp), \ens \De_\om\in\Der_\C(\tiRsp),
  \quad  \text{for $\om\in\C^*$}
  \edla
  as follows:
  given $\wt\ph=\gB\ii(\C\,\de+\wh\ph)\in\tiRsp$,
  Definition~\ref{defendlesscont} allows us to find a finite subset of
  $]0,\om[$, the points of which we denote by
  $\om_1\prec\cdots\prec\om_{r-1}$ (with reference to the natural
  total order on the line-segment), so that Definition~\ref{defDepDe}
  can be copied verbatim, except that the $\om_k$'s do not necessarily
  lie on any lattice
  (take any $L>\abs\om$ and $\{\om_1,\ldots,\om_{r-1}\}\defeq
  F_L\cap\,]0,\om[$ in Definition~\ref{defendlesscont}).
  
  Much of what has been said in the case of a lattice can be
  generalized to the case of~$\tiRsp$; in particular, this is a
  differential subalgebra of $(\C[[z\ii]],\frac{\dd\;}{\dd z})$
  and, in the summable case, we have a passage
  formula generalizing \eqref{eqdefDDpd}--\eqref{eqlogDDpd}:
  \begla
\gS^{\{\th\}} = \gS^{\{\th'\}} \circ \bigg( \sum_{\uom\in\uOmp}
P_{\th,\th'}^\uom \, \ee^{-\norm\uom z}\De_\uom \bigg),
\qquad \th<\th',
\edla
(at least in restriction to the simple resurgent series for which the
action of the \rhs\ is well-defined),
with a scalar mould~$P_{\th,\th'}$ defined by
\begla
P_{\th,\th'}^{\uom} \defeq
\begin{cases}
  1 & \text{if $r(\uom)=0$} \\[1ex]
  \frac{1}{r_1!\cdots r_s!} & \text{if
    $\th<\arg\om_1\le\cdots\le\arg\om_r<\th'$} \\[1ex]
  0 & \text{otherwise,}
  \end{cases}
\edla
where, in the second case, $\uom=\om_1 \cdots \om_r\neq\est$, $s$ counts the
number of pairwise distinct $\arg\om_j$'s, and $r_1,\ldots,r_s$ count
their multiplicities ($r_1+\cdots+r_s=r$).

\subsubsection{Relaxing the constraint on the nature of singularities} 

\begin{figure}[h]
  \caption{$\th$-rotated Hankel contour.}
\label{figHankCont}
\begin{center} 
  \includegraphics[width=.295\textwidth]{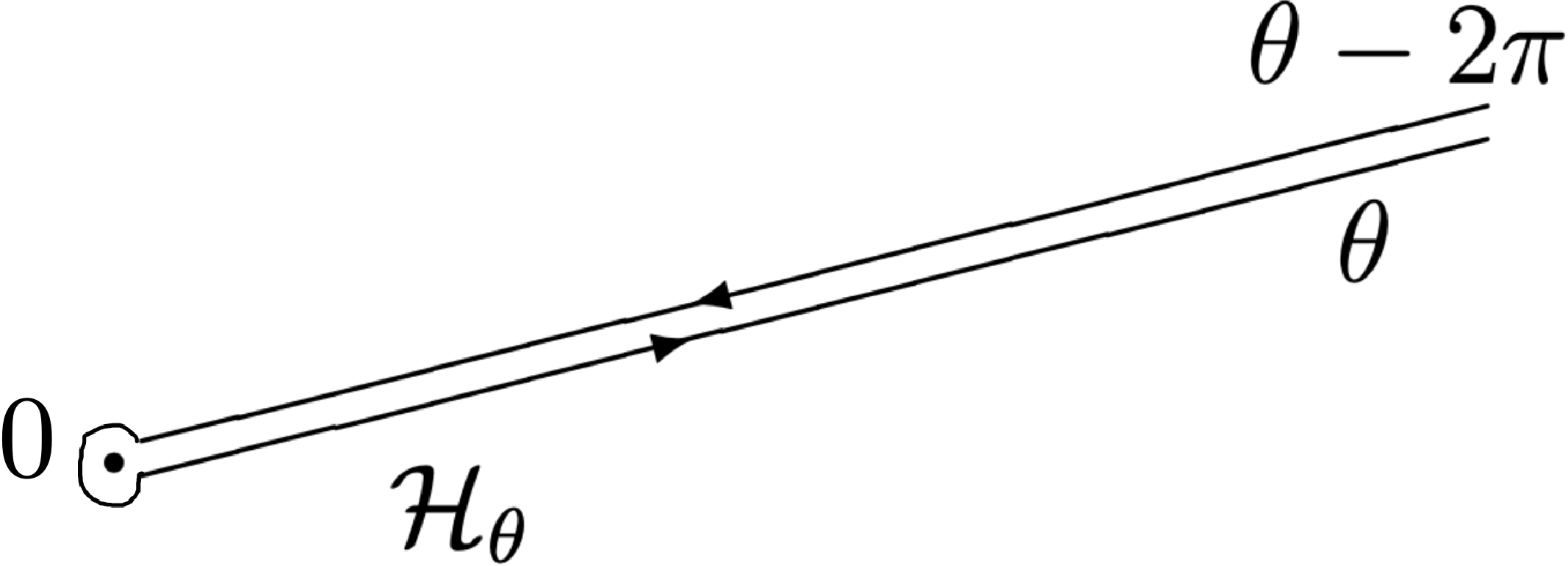}
\end{center}
\end{figure}

Our earlier computation of a difference of Laplace
transforms 
illustrated by Figure~\ref{figpathdiff} relied on the fact that,
given a direction $\th\in\R$, one has
\beglab{eqfirstHklLapl}
a_0 = \int_{\Hth} \ee^{-z\ze} \frac{a_0}{2\pi\I\ze}\,\dd\ze,
\ens
(\gL^\th\wh\chi)(z) = \int_{\Hth} \ee^{-z\ze} \,\chb\chi(\ze)\,\dd\ze,
\edla
where
\beglab{eqdefchb}
\chb\chi(\ze) \defeq \wh\chi(\ze) \, \frac{\log\ze}{2\pi\I}
+ F(\ze), \quad F(\ze)\in\C\{\ze\},
\edla
and $\Hth$ is a rotated Hankel contour as on
Figure~\ref{figHankCont}.
%

Let us thus introduce the Hankel-type Laplace
transform
\begla
(\chnLth\ch\psi)(z) \defeq
\int_{\Hth} \ee^{-z\ze} \,\ch\psi(\ze) \,\dd\ze.
\edla
A simple singularity at~$0$ of the form
\beglab{eqdefchpsisp}
\ch\psi(\ze) \defeq \frac{a_0}{2\pi\I\ze} +\chb\chi(\ze),
\edla
encoded by $a_0\de+\wh\chi(\ze)\in\C\,\de\oplus\C\{\ze\}$,
is mapped to a function that is asymptotic to a formal series
$a_0+\gB\ii\wh\chi \in \C[[z\ii]]$.
This gives a clue as to how we can deal with much more general
singularities, provided they are still isolated.

Let~$\wt\C$ denote the Riemann surface of the logarithm with its
points denoted by $\ze=r\,\ee^{\I\th}$, $r>0$, $\th\in\R$, \ie we view
it as a covering space over~$\C^*$ with $\ze\mapsto\ze\,\ee^{-2\pi\I}$
as a deck transformation.
We now define~$\ANA$ to be the space of all functions~$\ch\psi(\ze)$
analytic in a domain of the form $\{r<h(\th)\}$ for some
continuous positive function~$h$, \ie the totality of all possibly
singular germs
that are regular
in a spiralling neighbourhood of $\ze=0$.
Examples are singular germs with a simple singularity,
like~\eqref{eqdefchpsisp}, or higher order poles, or
essential singularities, or convergent expansions involving
non-integer powers of~$\ze$ and powers of~$\log\ze$
as in~\eqref{eqdefIJsig},
to name but a few.

\begin{Def}
  A \emph{singularity} is any member of $\SING\defeq
  \ANA/\C\{\ze\}$. We denote the canonical projection by
\begla
\ch\psi\in\ANA \mapsto \sing_0\bigl(\ch\psi(\ze)\bigr) \in \SING.
\edla
The \emph{minor} of a singularity $\tr\psi =
\sing_0\bigl(\ch\psi(\ze)\bigr)$ is
\begla
\wh\psi(\ze) \defeq \ch\psi(\ze) - \ch\psi(\ze\,\ee^{-2\pi\I}) \in \ANA.
\edla
We call \emph{resurgent singularity} any~$\tr\psi$ whose
minor~$\wh\psi(\ze)$ is endlessly continuable in the sense of Definition~\ref{defendlesscont}.
The space of all resurgent singularities is denoted by~$\trR$.
  \end{Def}

  The subspace $\trRspOm$ of~$\trR$ alluded to at the beginning of
  this article is precisely
  $\C\,\de\oplus\bem\big(\whRspOm)$, with the notation
  \begla
  \de \defeq \sing_0\!\Big( \frac{1}{2\pi\I\ze} \Big), \quad
  \bem\wh\chi \defeq \sing_0\!\Big( \wh\chi(\ze) \,
  \frac{\log\ze}{2\pi\I} \Big)
  \edla
  (from now on, we no longer view~$\de$ as a symbol but rather as a
  singularity).

It turns out that one can define a commutative convolution on~$\trR$
that extends the one inherited from $\whRspOm$ and makes it an
algebra, for which~$\de$ is the unit.
  
  Observe that the Hankel-type Laplace transform passes to the quotient:
  given a singularity whose minor has at most exponential growth at
  infinity in the direction~$\th$, we can set
  $\trnLth\tr\psi \defeq \chnLth\ch\psi$ for any
  representative~$\ch\psi$ of~$\tr\psi$.

However, the asymptotic expansions as $\abs z\to\infty$ of the
resulting functions can be much more general than mere power series.
The direct generalization of $\tiRspOm$ and~$\tiRsp$, the so-called
\emph{formal model of Resurgence}, $\tiR$, is thus more delicate to define
\parencite{Eca85}.
Suffice it to say that $\tiR$ is an algebra containing monomials like
$\wt I_\sig \defeq z^{-\sig}$ or
$\wt J_\sig \defeq -z^{-\sig}\log z$, to be viewed as inverse Borel
transforms of the singularities represented by
\beglab{eqdefIJsig}
\chn I_\sig \defeq g(\sig)\ze^{\sig-1},
\quad \chn J_\sig \defeq g(\sig)\ze^{\sig-1}\log\ze + g'(\sig)\ze^{\sig-1},
\edla
where $g(\sig) \defeq \frac{1}{2\pi\I} \ee^{\I\pi\sig}\Ga(1-\sig)$ for $\sig\in\C-\Zp$
(the case $\sig\in\Zp$ must be treated separately),
as well as $z^{-\sig}(\log z)^m$ for any $m\in\Zp$.

Accordingly, in practice, one often encounters resurgent transseries
of the form
\begla
\sum_{k\ge0} \ee^{-\om_k z} \wt\psi_k(z),
\quad \wt\psi_k(z)\in z^{-\sig_k}\C[[z\ii]]\cap\tiR
\edla
with a certain sequence of exponents $(\sig_k)$.
This happens \eg with one-parameter transseries solutions to Painlev\'e equations,  
in which case there are also two-parameter transseries containing logarithms
\parencite{BSSV,vSV}.

The general alien operators 
  \begla
  \De^+_\om \in \End_\C(\trR), \ens \De_\om\in\Der_\C(\trR),
  \quad  \text{for $\om\in\wt\C$}
  \edla
  are defined by the natural generalization of what was done in~$\tiRsp$, \eg $\De_\om\tr\psi\defeq$
  \begla
  \sum_{\eps\in\{+,-\}^{r-1}}
  \frac{p(\eps)!q(\eps)!}{r!}
  \sing_0\!\big((\cont_{\ga(\eps)} \wh\psi)(\om+\ze)\big),
  \edla
  where $\om_1\prec\cdots\prec\om_{r-1}$ are the points of $]0,\om[$
  (now in~$\wt\C$) to be circumvented when following the analytic
  continuation of the minor.
  
\subsubsection{Examples from mathematical physics}
In \parencite{GK_Fadd}, the logarithm of Faddeev's quantum dilogarithm $\Phi_\bb(x)$
(a special function that plays a key role in quantum Teichm\"uller
theory and Chern-Simons theory) is shown to arise
from the Borel-Laplace summation of a simple resurgent series:
for fixed $x\in\C$ with $\arg x\in(-\pi,\pi)$,
\begla
\log\Phi_{\bb}(\tfrac{x}{2\pi\bb}) =
\tfrac{z}{2\pi\I} \operatorname{Li}_2(-\ee^x) +
(\gS^{\{0\}}\wt G)(z,x)
\edla
for $z=\bb^{-2}$, where $\wt G(z,x)\in z\ii\C[[z^{-2}]]$ has
meromorphic Borel transform with simple poles that lie on a countable
union of lines passing through~$0$. More precisely, $\wh G(\ze,x)$ has
a simple pole with residue $\frac{(-1)^n}{2\pi\I n}$ at
\begla
\om_{m,n}(x) \defeq n\big( x + (2m+1)\I\pi\big), \quad
m\in\Z,\; n\in\Z^*.
\edla
Since resurgent series are stable under nonlinear operations like
exponentiation \parencite{Eca85} (details can be found in
\parencite{KS20}), it follows that
\begla
\Phi_{\bb}(\tfrac{x}{2\pi\bb}) =
\exp\!\big({\tfrac{z}{2\pi\I} \operatorname{Li}_2(-\ee^x)}\big)\,
(\gS^{\{0\}}\wt H)(z,x)
\edla
with $\wt H\defeq \exp \wt G \in\tiRsp$. The Borel transform~$\wh H(\ze,x)$ is
not meromorphic, but its branches may be singular only for
$\ze\in x\Z+\I\pi\Z$ (because convolution and a fortiori the Borel
counterpart of exponentiation tend to add singular points) and alien
chain rule gives
\begla
\De_{\om_{m,n}(x)} \wt H(z,x) = \frac{(-1)^n}{n}\wt H(z,x),
\edla
which indicates superpositions of simple poles and logarithmic
singularities.
Notice that $\De_\om\wt H=0$ if $\om\in x\Z+\I\pi\Z$ is not among the
$\om_{m,n}(x)$'s, but some branch of~$\wh H$ (though not the principal one) may
be singular at such a point~$\om$.

Borel transforms displaying singularities arranged on a countable
union of lines through~$0$, like a peacock pattern, have been observed
\begin{itemize}
\item
  in \parencite{PS} and \parencite{CSPMMRS}, in the context of topological string theory
  (studying a resurgent transseries for the free energy of topological
  strings on the resolved conifold and the local $\PPr^2$ toric Calabi-Yau threefold),
\item
in \parencite{GGM_peacock}, in the context of Chern-Simons theory
(studying the partition function of the complement of a hyperbolic
knot in~$\SSS^3$).
\end{itemize}


\subsection{Mould Calculus}   \label{secMC}


\subsubsection{The mould algebra}


We have encountered $\tiRspOm$-valued moulds in the context of alien
calculus, and~$\Om'$, a rank-one lattice minus the origin, was then used as alphabet.
We now develop the basics of mould calculus in the broader context of
a commutative $\Q$-algebra~$A$ and an arbitrary alphabet~$\fN$.
We denote by~$\uN$ the free monoid on~$\fN$,
which is not supposed to be countable or contained in~$\C$, but, when
necessary, we assume~$\fN$ to be a commutative semigroup, in which
case we use the notation $\norm\uom \defeq \om_1+\cdots+\om_r$ for any
$\uom=\om_1\cdots\om_r \in \uN$.

An \emph{$A$-valued mould on~$\fN$} is just a map $M\col \uN \to A$;
it is customary to use the notation~$M^\uom$ instead of
$M(\uom)$. Word concatenation in~$\uN$ induces mould multiplication:
\begla
(M\times N)^\uom \defeq \sum\limits_{(\ua,\ub) \ \text{such that} \
  \uom=\ua\,\ub}M^{\ua}N^{\ub},
\edla
and the space $A^\uN$ of all moulds is thus an associative
$A$-algebra, non-commutative if~$\fN$ has more than one element, whose
unit is the mould~$\idm$ defined by
$\idm^\est = 1$ and $\idm^\uom = 0$ for $\uom\neq\est$
($A^\uN$ is sometimes denoted by $A\lan\!\lan \fN \ran\!\ran$ in
algebraic combinatorics).
A mould~$M$ is invertible if and only if~$M^\est$ is invertible in~$A$.

There is also an associative non-commutative \emph{mould composition}:
for each $U\in A^\uN$, we have an $A$-algebra homomorphism
$M\mapsto M \circ U$ defined by
$(M \circ U)^\est\defeq M^\est$ and,
for $\uom\neq\est$,
\begla
(M \circ U)^\uom\defeq \sum_{ \substack{s\ge1,\,
\uom^1,\dotsc,\uom^s\neq\est  \\
\uom = \uom^1 \cdots \uom^s}  }
M^{\norm{\uom^1} \cdots \norm{\uom^s}}
U^{\om^1} \cdots U^{\om^s}.
\edla
We denote by $r(\uom)$ the length of a word~$\uom$.
The identity mould~$I$ defined by $I^\uom\defeq 1$ if $r(\uom)=1$
and~$0$ otherwise satifies $M\circ I=M$ and $I\circ U=U$ for all
moulds~$M$ and~$U$ such that $U^\est=0$.
Such~$U$ has a composition inverse if and only $U^\uom$ is invertible
in~$A$ whenever $r(\uom)=1$.

A mould~$M$ is said to have order $\ge p$ if $M^\uom=0$ whenever $r(\uom)<p$.
There is correspondingly a notion of formal convergence (a series of
moulds $\sum M_k$ is convergent if $\ord(M_k)\to\infty$, so that for
any given~$\uom$ only finitely many terms contribute), which for
instance allows to define mutually inverse bijections 
\beglab{eqdefexplog}
\{\, M\in A^\uN \mid M^\est=0 \,\}
\quad \overset{\exp}{\underset{\log}{\rightleftarrows}} \quad
\{\, M\in A^\uN \mid M^\est=1 \,\}
\edla
by the usual exponential and logarithm series.


\subsubsection{Mould expansions and mould symmetries}


$A$-valued moulds often provide the coefficients of certain multi-indexed
expansions in an associative $A$-algebra~$\cB$.
We may then restrict to finite-support moulds so as to avoid infinite
expansions, as we did earlier with $\fN=\Om'$, $A=\tiRspOm$ and
$\cB=\End_\C(A)$ in Section~\ref{secFree}, or we may assume~$\cB$ to be a complete filtered
associative algebra, thus endowed with a notion of formal convergence.

Then, given a family $(B_\om)_{\uom\in\fN}$ in~$\cB$, we extend it
  to~$\uN$ by defining $\BB_\est\defeq 1_\cB$ and
  \begla
  \BB_\uom \defeq B_{\om_r}\cdots B_{\om_1}
  \ens\text{for any $\uom=\om_1\cdots\om_r\in\uN$,}
  \edla
  and consider the \emph{mould expansions}
  \begla
  M\BB \defeq \sum_{\uom\in\uN} M^\uom \,\BB_\uom \in \cB
  \edla
associated with moulds $M\in A^\uN$.
One can check that $(M\times N)\BB = (N\BB)(M\BB)$,
$(\exp M)\BB = \exp(M\BB)$, $(\log M)\BB = \log(M\BB)$
when these mould expansions make sense.

If $\cB$ is only supposed to be a Lie algebra, one can define another
kind of mould expansion:
we then set $\BB_\cuom\defeq
[B_{\om_r},[\ldots[B_{\om_{2}},B_{\om_1}]\ldots]]$ and
\begla
M[\BB] \defeq \sum_{\uom\neq\est} \frac{1}{r(\uom)}M^\uom \BB_\cuom.
\edla
Then, recalling the definition~\eqref{eqaltPhi} of alternality,
\begla
\text{$M$ and~$N$ alternal} \imp
[M,N][\BB] = \big[ M[\BB], N[\BB] \big].
\edla
If~$\cB$ is an associative algebra with commutator as Lie bracket, the
two kinds of alternal mould expansions coincide,
as in~\eqref{eqPhiDDalt}---\cf the classical Dynkin-Specht-Wever
projection lemma
\parencite{CR93}.

The set of alternal moulds~$U$ that have a composition inverse is a
group for mould composition.

\noindent\emph{Symmetrality and alternality.}
Parallel to the definition~\eqref{eqaltPhi} of alternality, we have
\begin{Def}
  A mould~$M$ is called \emph{symmetral} if
  \beglab{eqsymPhi}
    M^\est=1_A \ens\text{and}\ens
    \sum_{\uom\in\uN} \sh{\ua}{\ub}{\uom} M^{\uom} = M^\ua M^\ub
    \edla
    for any $\ua,\ub\in\uN$.
\end{Def}

Symmetral moulds form a group for mould multiplication; the logarithm
map~\eqref{eqdefexplog} maps it bijectively to the space of alternal
moulds, which is a Lie algebra for mould commutator.
Mould expansions enjoy special properties when each~$B_\om$ acts as a
derivation on some auxiliary algebra, as is the case of $\De_\om \in
\Der_\C(A)$ when $A=\tiRspOm$:
\begin{align*}
\text{$M$ alternal} &\imp
                      \text{$M\BB$ is a derivation}\\
\text{$M$ symmetral} &\imp
                      \text{$M\BB$ is an automorphism.}
\end{align*}
The concepts of symmetrality and alternality, introduced in
\parencite{Eca81}, are related to certain combinatorial Hopf algebras,
as emphasized by \textcite{Me09} in his work on the renormalization theory
in perturbative quantum field theory.
See \parencite{LSS19} for the relation to the {Baker}-{Campbell}-{Hausdorff} formula.

\noindent\emph{Other types of symmetries.}
One may also consider mould expansions based on a family
$(B_\om)_{\uom\in\fN}$ of operators that,
instead of being derivations,
satisfy the same modified Leibniz rule~\eqref{eqmodifLeibniz} as the
$\De^+_\om$.
We then have parallel statements for the mould expansions associated
with moulds enjoying
`symmetr\hspace{-.075em}\emph{e}lity' or `alter\-n\hspace{-.055em}\emph{e}lity',
two notions defined in a manner
analogous to symmetrality and alternality but involving,
instead of shuffling,
`contracting shuffling' (also called stuffling).

\textcite{EcaTale} also introduced
`symmetr\hspace{-.065em}\emph{i}lity' and
{`altern\hspace{-.045em}\emph{i}lity'} in his works on MZVs, as well
as many other structures in the context of `bimoulds'---see below.


\subsubsection{The hyperlogarithmic mould}


We now come to the precise statement of the result referred to in
Section~\ref{secFree}, of which Theorem~\ref{thmfreeness} was a simple consequence:
\begin{thm}[\parencite{Eca81}]   \label{thmexistgUreinf}
 Let $A\defeq\tiR$ (resp.~$\tiRsp$, resp.~$\tiRspOm$) and $\fN\defeq\wt\C$
 (resp.~$\C^*$, resp.~$\Om'$).
 There exists a symmetral $A$-valued mould
     $(\gU^{\uom})_{\uom\in\uN}$ such that, for every
    $\om_0\in\fN$ and $\uom=\om_1\cdots\om_r\in\uN$,
    \begla
    \De_{\om_0}\gU^{\om_1\cdots\om_r} = 
      \begin{cases}
        \gU^{\om_2\cdots\om_r} & \text{if $r\ge1$ and $\om_0=\om_1$}
        \\[1ex]
        0 & \text{otherwise}.
      \end{cases}
      \edla
\end{thm}
\begin{proof}[Sketch of proof with $A=\tiRsp$ and $\fN=\C^*$]
  Pick an entire function~$\wh a_\om$ such
  that $\wh a_\om(\om)\neq0$ for each $\om\in\fN$.
The equations $\tcVae(z)\defeq 1$ and, for $\uom\neq\est$,
\begla
\big(\tfrac{\dd\;}{\dd z} + \norm\uom\big) \tcVao(z)
= - \tcVaop(z)(\gB\ii \wh a_{\om_r})(z)
\edla
inductively define $\tcVao(z)\in z\ii\C[[z\ii]]$, with Borel transforms
\beglab{eqBtrsftcV}
\hcVaun(\ze) = \frac{\wh a_{\om_1}(\ze)}{\ze-\om_1},
\qquad
\hcVao(\ze) = \frac{\hcVaop * \wh a_{\om_r}}{\ze-\norm{\uom}}.
\edla
The resulting mould~$\tcVab$ is $A$-valued and symmetral.

Elementary manipulations show that, for each $\eta\in\fN$, there
is an alternal scalar mould $V(\eta)$ such that
\beglab{eqaldertcV}
\De_\eta\tcVab = -V(\eta) \times  \tcVab,
\qquad 
V^\uom(\eta) \neq 0 \imp \norm{\uom}=\eta.
\edla
By~\eqref{eqBtrsftcV}, 
$V^{\om_1}(\om_1) = -2\pi\I\, \wh a_{\om_1}(\om_1)$.

We define an alternal mould by $V = \sum_{\eta\in\fN}  V(\eta)$.
It has a composition inverse, that we denote by~$-U$.
Elementary manipulations show that
$\gU \defeq \tcVab \circ U$ satisfies the requirements of Theorem~\ref{thmexistgUreinf}.
Details can be found in \parencite{Eca81,mouldSN},
where the resurgent series~$\gU^{\uom}(z)$ are called `$\De$-friendly resurgence
monomials', in contrast with the `$\pa$-friendly resurgence monomials'~$\tcVao(z)$.
\end{proof}

The \emph{hyperlogarithmic mould} is the scalar mould~$U$ obtained
when $\fN=\Z^*$ and $\wh a_\om(\ze)=1$ for all~$\ze$ and~$\om$.
In that case, there are scalar moulds $L(\eta)$
($\eta\in\Z^*$) such that
\beglab{eqDepcVL}
\De^+_\eta\tcVab = L(\eta) \times  \tcVab,
\qquad 
L^\uom(\eta) \neq 0 \imp \norm{\uom}=\eta,
\edla
giving rise to a `multiple logarithm' mould
\beglab{eqdefLbul}
L^\uom = \sum_\eta L^\uom(\eta) =
\int_{\Ga^+}
\frac{2\pi\I \,\dd\ze_{1}\dotsm\dd\ze_{r-1}}{%
(\ze_{1}-\wc\om_{1})\dotsm(\ze_{r-1}-\wc\om_{r-1})},
\edla
where $\wc\om_1=\om_1, \wc\om_2=\om_1+\om_2, \dotsc$ and~$\Ga^+$
connects~$0$ and $\wc\om_r=\om_1+\cdots+\om_r$ by circumventing
integers to the right.
The moulds $L_{\pm} \defeq \idm+\sum_{\pm\eta>0}L^\uom(\eta)$ are symmetral.
These moulds are related to the MZVs.


\subsubsection{Mould aspects of the MZV world}


\textcite[\S12e]{Eca81} introduced the multizetas
\begla
\ze(s_1,s_2,\dotsc,s_r) = \sum_{n_1>n_2>\dotsb>n_r>0}
\frac{1}{n_1^{s_1}n_2^{s_2}\dotsm n_r^{s_r}}
\edla
as a scalar mould on~$\Zp$. He later studied systematically their
natural generalization, known as \emph{coloured} (or \emph{modulated})
\emph{multizeta values} (`MZV'),
\begla
\ZE^{\left(\begin{smallmatrix}
\eps_1,&\dotsc\,,&\eps_r\\s_1,&\dotsc\,,&s_r\end{smallmatrix}\right)}
= \sum_{n_1>\dotsb>n_r>0}
\frac{\ee^{2\pi\I(n_1\eps_1+\dotsb+n_r\eps_r)}}{n_1^{s_1}\dotsm n_r^{s_r}}
\edla
for $s_1,\dotsc,s_r\in\Zp$, $\eps_1,\dotsc,\eps_r\in\Q/\Z$
(with suitable convention to handle possible divergences).
The mould~$\ZE$ is symmetrel.
It is called a {\em bimould} because the letters of the alphabet are naturally given
as members of a product space, here $\Zp\times(\Q/\Z)$.

There is a related mould on $\{0\}\cup \,\exp(2\pi\I\Q)$, namely
$\WA^{\al_1,\dots,\al_\ell} \defeq$
\begla
(-1)^{\ell_0} \int_{0<\ze_1<\dotsb<\ze_\ell<1}
\frac{\dd \ze_1\dotsm\dd \ze_\ell}{(\al_1-\ze_1)\dotsm(\al_\ell-\ze_\ell)},
\edla
where $\ell_0$ is the number
of~$0$'s among the~$\al_j$'s.
With a suitable extension of this definition when $\al_1=0$ or
$\al_\ell=1$, the resulting mould is symmetral and related to the multiple logarithm mould~$L$:
\begla
\WA^{\al_1,\dotsc,\al_\ell} = \frac{1}{2\pi\I}(-1)^{\ell-\ell_0}
L^{\al_1,\al_2-\al_1, \dotsc, \al_\ell-\al_{\ell-1}, 1-\al_\ell}
\edla
(at least if $\al_j\in\{-1,0,1\}$),
and 
\begla
\ZE^{\left(\begin{smallmatrix}
\eps_1,&\dotsc\,,&\eps_r\\s_1,&\dotsc\,,&s_r\end{smallmatrix}\right)}
= \WA^{\wh e_r, 0^{[s_r-1]}, \dotsc, \wh e_1, 0^{[s_1-1]}},
\edla
with $\wh e_j = \ee^{2\pi\I(\eps_1+\dotsb+\eps_j)}$.

Dealing with bimoulds makes it possible
to define a host of new operations and structures;
this is the starting point of a whole theory, aimed at describing the algebraic
structures underlying the relations between the multizeta values.
A few references are
\parencite{Wald2000}, \parencite{EcaTale,EcaBil,EcaScramble}, \parencite{LS20}, \parencite{FK22}.


\subsection{Applications}



Resurgence Theory originated with local analytic dynamics
\parencite{Eca81,Eca85}. Indeed, the Abel equation that governs the
dynamics of tangent-to-identity holomorphic germs in $(\C,0)$ belongs
to a class of difference equations giving rise to resurgent series.
A related problem is that of nonlinear ODEs of saddle-node type
\parencite{MR82}---see also \parencite{SK22}.

In such problems, the resurgent analysis leads to a `Bridge equation',
according to which the action of the alien derivations on the various series of the
problem amounts to the action of certain ordinary differential
operators (this self-reproduction phenomenon is the reason why
\'Ecalle chose the name `resurgence');
the formulas involve coefficients (usually called Stokes coefficients)
that can be used to describe the moduli space of the problem.
See \parencite{mouldSN} for a detailed analysis in the saddle-node
case, where the solutions are constructed as mould expansions
involving a resurgent symmetral mould akin to the~$\tcVab(z)$ of the
proof of Theorem~\ref{thmexistgUreinf} and operators~$B_\om$ that are
essentially the homogeneous components of the vector field associated
with the ODE.

In fact, by its ability to handle multiply indexed series of
operators, mould calculus proves to be a flexible tool to construct
formal solutions to dynamical problems.
In some cases, one can even reach analytic conclusions by means of the
\'Ecalle's `arborification' technique---see \parencite{FM17} for
connections between arborified moulds and the Connes-Kreimer Hopf
algebras of trees and graphs and their applications to perturbative
quantum field theory.
Other examples of application of mould calculus to formal or analytic
dynamical problems can be found in \parencite{Eca92, EV98},
\parencite{PSRCD}, \parencite{NPSTLMP}, \parencite{FMS18}.


Already in the early 1980s, a different source of resurgence was
identified by \textcite{VorosBour,Vor83} in his seminal work on the
exact WKB method and the Borel transform of the Jost function.
One may argue that the idea of resurgence in physics was somehow
implicit in the work of \textcite{Par78}, 't~Hooft
\parencite{tHooft79} and others
(C.~M.~Bender,
\'E.~Br\'ezin,
C.~Itzykson,
L.~N.~Lipatov,
A.~Voros,
T.~T.~Wu,
J.~Zinn-Justin,
J.-B.~Zuber,
to name but a few).
The encounter between this circle of ideas and \'Ecalle's theory
triggered a new line of research, pursued notably by F.~Pham and
co-authors in various papers, \eg \parencite{DDP93,DP99},
that recently led to new developments in Quantum Field Theory since
the advent of spectral networks \parencite{GMN},
\parencite{KS22}---see \parencite{primer}.

Resurgence is now more and more used in Topological Quantum Field
Theory, Quantum Modularity \parencite{GZ}, Chern-Simons theory
\parencite{GMP,3dmod,AM22,HLSS}, deformation quantization
\parencite{Garayetal, LSSMoyal},
as well as in studies of (Ward-)Schwinger-Dyson equations and
renormalisation group
\parencite{BC18, 
  BB,
  BRphi,
  BD}.




\subsubsection*{Acknowledgements}

The author thanks Capital Normal University (Beijing) for its
hospitality.
This paper is partly a result of the ERC-SyG project, Recursive and
Exact New Quantum Theory (ReNewQuantum) which received funding from
the European Research Council (ERC) under the European Union's Horizon
2020 research and innovation programme under grant agreement No
810573.
This work has been partially supported by the project
CARPLO of the Agence Nationale de la recherche (ANR-20-CE40-0007).

\printbibliography

\end{document}